\documentclass[prb,preprint,showpacs]{revtex4}

\usepackage{amsmath,amssymb}

\usepackage{graphicx}

\begin{document}

\widetext

\title{Quantized Lattice Dynamic Effects on the Spin-Peierls Transition}

\author{Christopher J. Pearson$^{1}$\footnote{E.mail address: chris.pearson@chem.ox.ac.uk
}, William Barford$^{1}$\footnote{E.mail address:
william.barford@chem.ox.ac.uk }, and Robert J. Bursill$^{2}$}

\affiliation{$^1$Department of Chemistry, Physical and Theoretical
Chemistry Laboratory, University of Oxford, Oxford, OX1 3QZ, United
Kingdom\\
$^2$School of Physics, University of New South Wales, Sydney, New
South Wales 2052, Australia}

\begin{abstract}
The density matrix renormalization group  method is used to
investigate the spin-Peierls transition for Heisenberg spins coupled
to quantized phonons. We use a phonon spectrum that interpolates
between a gapped, dispersionless (Einstein) limit to a gapless,
dispersive (Debye) limit. A variety of theoretical probes are used
to determine the quantum phase transition, including energy gap
crossing, a finite size scaling analysis, bond order
auto-correlation functions, and bipartite quantum entanglement. All
these probes indicate that in the antiadiabatic phonon limit a
quantum phase transition of the Berezinskii-Kosterlitz-Thouless type
is observed at a non-zero spin-phonon coupling, $g_{\text c}$.  An
extrapolation from the Einstein limit to the Debye limit is
accompanied by an increase in $g_{\text c}$ for a fixed optical
($q=\pi $) phonon gap. We therefore conclude that the dimerized
ground state is more unstable with respect to Debye phonons, with
the introduction of phonon dispersion renormalizing the effective
spin-lattice coupling for the Peierls-active mode. We also show that
the staggered spin-spin and phonon displacement order parameters are
unreliable means of determining the transition.
\end{abstract}


\maketitle

\section{Introduction}

Since the discovery of high-temperature superconductivity in doped
antiferromagnets there has been a marked increase in interest --
both theoretical and experimental -- in low-dimensional quantum
magnetism. However, the effect of the interaction of quantum spins
with further degrees of freedom such as disorder, phonons, and holes
produced by doping remains relatively poorly understood.

The instability of the spin-$\frac{1}{2}$ antiferromagnetic
Heisenberg chain to a static uniform distortion gives rise to the
so-called spin-Peierls (SP) transition.   This occurs because the
explicit dimerization opens a gap, $\Delta$, in the spin-excitation
spectrum, lowering the total magnetic energy by an amount that
offsets the accompanying increase in lattice energy.

The SP instability is itself the antiferromagnetic analogue of the
  Peierls transition\cite{peierls}: a half-filled
one-dimensional  metallic phase is unstable with respect to a
commensurate periodic lattice distortion of wave vector $q=2k_F$.
Indeed, SP models represent the large on-site coupling limit of the
corresponding half-filled Hubbard-Peierls Hamiltonian: for infinite
inter-electron repulsion, charge degrees of freedom are effectively
quenched resulting in the loss of electron itinerancy for a
half-filled band.  For linear chains under open boundary conditions
the Jordan-Wigner transformation maps the Heisenberg-SP model onto a
spinless fermion-Peierls model with nearest-neighbour repulsion,
highlighting the decoupling of spin and charge degrees of freedom
and explaining the SP nomenclature.

The SP instability is well understood in the \emph{static-lattice}
limit for which the frequency,  $\omega_\pi$, of the Peierls-active
mode is taken to be much smaller than the antiferromagnetic exchange
integral, $J$.  In this adiabatic phonon limit the ground state (GS)
is known to have a broken-symmetry staggered dimerization for
arbitrary electron-phonon (e-ph) coupling.  Experimentally, such
behavior was first observed in the 1970s for the organic compounds
of the TTF and TCNQ series\cite{ttf}.  For many
quasi-one-dimensional materials, however, the zero-point
fluctuations of the (quantized) phonon field are comparable to the
amplitude of the Peierls distortion \cite{mc, wellein2, lav}.  In
CuGeO$_3$ \cite{hase}, for example, Cu$^{2+}$ ions form well
separated spin-$\frac{1}{2}$ chains with an exchange interaction
that couples to high-frequency optical phonons
$\omega_\pi\sim\mathcal{O}(J)$. CuGeO$_3$ has since become a
paradigm of inorganic antiadiabatic SP behaviour, stimulating
several numerical studies of dynamical phonon
models\cite{augier,wellein,bursill}.

Using both the density-matrix renormalization group (DMRG)
method\cite{bursill} and  renormalization group (RG) methods within
a bosonization scheme \cite{citro}, it has been demonstrated that
quantum fluctuations destroy the Peierls state for small, non-zero
couplings in both the spinless and spin-$\frac{1}{2}$ Holstein
models at half-filling. Analogous results for the $XY$-SP model with
gapped, dispersionless (Einstein) phonons were also obtained by
Caron and Moukouri \cite{caron}, using finite-size scaling analysis
of the spin gap to demonstrate a power-law relating the critical
coupling and the Peierls-active phonon frequency: $g_{\text
c}^{XY}\sim\omega_\pi^{0.7}$.  For models with sufficiently large
Einstein frequency, gapped phonon degrees of freedom can be
integrated away to generate a low-energy effective-fermion
Hamiltonian characterized by instantaneous, non-local
interactions\cite{kuboki}.  For spinless models, RG equations
indicate that unless the non-local contribution to the umklapp term
has both the right sign and a bare (initial) value larger than a
certain threshold, the umklapp processes are irrelevant and the
quantum system is gapless \cite{bourbon}. Conversely, if the
threshold condition is satisfied, the umklapp processes and vertex
function grow to infinity, signalling the onset of gapped
excitations and a dimerized lattice.

For the Su-Schrieffer-Heeger (SSH) model, Fradkin and Hirsch
undertook an extensive  study of spin-$\frac{1}{2}$ ($n=2$) and
spinless ($n=1$) fermions using world-line Monte Carlo
simulations\cite{fradkin}.  In the antiadiabatic limit (i.e.\
vanishing ionic mass $M$), they mapped the system onto an
$n$-component Gross-Neveu model, known to exhibit long-ranged
dimerization for arbitrary coupling for $n\ge 2$ (although not for
$n=1$).  For $M>0$ an RG analysis shows the low-energy behavior of
the $n=2$ model to be governed by the zero-mass limit of the theory,
indicating that the spinful model presents a dimerized GS for
arbitrarily weak e-ph couplings.  The spinless model, on the other
hand, has a disordered phase for small coupling if $M$ is finite,
with an ordered phase realized for bare coupling in excess of a
certain threshold.  As $M\to \infty$ the size of the disordered
region shrinks to zero, reconnecting with the adiabatic result of
Peierls and Fr\"olich\cite{peierls}.  Later work by Zimanyi \emph{et
al.}\cite{zim} on one-dimensional models with both electron-electron
(e-e) and e-ph interactions showed they were found to develop a spin
gap if the combined backscattering amplitude $g_1^{\text
T}=g_1(\omega)+\tilde{g}_1(\omega)<0$, where $g_1(\omega)$ is the
contribution from electron-electron (e-e) interactions and
$\tilde{g}_1(\omega)<0$ is the e-ph contribution in the notation of
\cite{zim}.  Hence, for the pure \emph{spinful} SSH model, $g_1=0$
and $g_1^{\text T}<0$ for any nonzero e-ph coupling, implying a
Peierls GS for arbitrary e-ph coupling, in agreement with the
earlier MC results\cite{fradkin}.

In this article we study the influence of gapless, dispersive
antiadiabatic phonons  on the GS of the Heisenberg-Peierls chain.
That this model is yet to receive the same level of attention as its
gapful, dispersionless counterpart is due in part to the presence of
hydrodynamic modes, resulting in logarithmically increasing
vibrational amplitudes with chain length. To this end, the authors
of\cite{fradkin} and\cite{zim} assumed acoustic phonons to decouple
from the low-energy spin states involved in the SP instability,
motivating the retention of only the optical phonons close to
$q=\pi$.  In this regard, optical phonons have been expected to be
equivalent to fully quantum mechanical SSH phonons. For \emph{pure
Einstein phonons}, Wellein, Fehske, and Kampf\cite{wellein},
however, found that the singlet-triplet excitation is strongly
renormalized when phonons of all wavenumber are taken into account,
the restriction to solely the $q=\pi$ modes leading to a substantial
overestimation of the spin gap. Physically, this implies that the
spin-triplet excitation is accompanied by a local distortion of the
lattice, necessitating a multiphonon mode treatment of the lattice
degrees of freedom.  We anticipate, then, that truncating the
Debye-phonon spectrum to leave only those modes which couple
directly to the SP phase may be not be physically reasonable.

In this work we use the DMRG technique to numerically solve the
Heisenberg-Peierls model with a generalized gapped, dispersive
phonon spectrum. The phonon spectrum interpolates between a gapped,
dispersionless (Einstein) limit and a gapless, dispersive (Debye)
limit. We proceed by considering a system of Heisenberg spins
dressed with pure Einstein phonons for which we observe a
Berezinskii-Kosterlitz-Thouless (BKT) quantum phase transition at a
non-zero spin-lattice coupling. Progressively increasing the Debye
character of the phonon dispersion (at given phonon adiabaticity)
results in an increase in the critical value of the spin-lattice
coupling, with the transition remaining in the BKT universality
class (see Section \ref{bkt}).  These findings are corroborated by
an array of independent verifications: energy-gap crossings in the
spin-excitation spectra (see Section \ref{gapc}), finite-size
scaling of the spin-gap (see Section \ref{fssg}), bond order
auto-correlation functions (see Section \ref{dop}), and quantum
bipartite entanglement (see Section \ref{bipent}).

We note that earlier DMRG investigations of the Heisenberg-SP
Hamiltonian with Debye phonons  indicated a dimerized GS for
arbitrary coupling\cite{barford1}.  This conclusion was based on the
behavior of the staggered phonon order parameter, $m_{\text p}$,
(defined in Section \ref{dop}). In this paper we show that $m_{\text
p}$ is an unreliable signature of the transition.

In the next Section we describe the model, before discussing our
results in Section \ref{Se:3}.

\section{The Model}
The Heisenberg spin-Peierls Hamiltonian is defined by,
\begin{equation}\label{ham}
H=H_{\text{s-p}} + H_{\text{p}}.
\end{equation}
$H_{\text{s-p}}$ describes the spin degrees of freedom and the
spin-phonon coupling,
\begin{equation}\label{spph}
H_{\text{s-p}}=\sum_l[J+\alpha(u_{l+1}-u_l)]\textbf{S}_l\cdot
\textbf{S}_{l+1},
\end{equation}
where $\textbf{S}_l$ is the Pauli spin operator, $u_l$ is the
displacement of the $l$th ion from equilibrium, and $\alpha$ is the
spin-phonon coupling parameter.

$H_{\text{p}}$ describes the lattice degrees of freedom. In the Einstein
model the ions are decoupled,
\begin{equation}\label{ein}
H^E_{\text{p}}=\sum_l\frac{P_l^2}{2M}+\frac{1}{2}K\sum_lu_l^2.
\end{equation}
In the Debye model, however, the ions are coupled to nearest
neighbors,
\begin{equation}\label{deb}
H^D_{\text{p}}=\sum_l\frac{P_l^2}{2M}+\frac{1}{2}K \sum_l (u_{l+1} -
u_l)^2.
\end{equation}

For the Einstein phonons it is convenient to introduce phonon
creation, $b_l^\dagger$, and annihilation operators, $b_l$, for the
$l$th site via,
\begin{equation}\label{ul}
u_l = \left(\frac{\hbar}{2M\omega_X}\right)^{1/2}(b_l^\dagger + b_l)
\end{equation}
and
\begin{equation}\label{pl}
P_l = i\left(\frac{M\hbar\omega_X}{2}\right)^{1/2}(b_l^\dagger -
b_l),
\end{equation}
where
\begin{equation}\label{}
\omega_X = \omega_E = \sqrt{K/M} \equiv {\omega_b}.
\end{equation}
Making these substitutions in Eq. (\ref{spph}) and Eq. (\ref{ein})
gives,
\begin{equation}
H_{\text{s-p}}=J\sum_l\left[1+g_E\left(\frac{\hbar\omega_E}{J}
\right)^{1/2} (B_l-B_{l+1})\right]\textbf{S}_l\cdot \textbf{S}_{l+1}
\end{equation}
and
\begin{equation}
H^E_{\text{p}}=\hbar\omega_E\sum_l\left(b_l^\dagger
b_l+\frac{1}{2}\right),
\end{equation}
where $B_l=\frac{1}{2}(b_l^\dagger+b_l)$ is the dimensionless phonon
displacement and,
\begin{equation}\label{Eq:11}
g_E = \alpha\left(\frac{2}{M\omega_E^2J}\right)^{1/2} =
\alpha\left(\frac{2}{KJ}\right)^{1/2},
\end{equation}
is the dimensionless spin-phonon coupling parameter.

For the Debye phonons we introduce phonon creation and annihilation
operators defined by Eq. (\ref{ul})  and Eq. (\ref{pl}) where
\begin{equation}\label{}
     \omega_X = \omega_D = \sqrt{2K/M} \equiv \sqrt{2} {\omega_b}.
\end{equation}
Making these substitutions in Eq. (\ref{spph}) and Eq. (\ref{deb})
gives,
\begin{equation}
H_{\text{s-p}}=J\sum_l\left[1+g_D\left(\frac{\hbar\omega_D}{J}
\right)^{1/2} (B_l-B_{l+1})\right]\textbf{S}_l\cdot \textbf{S}_{l+1}
\end{equation}
and
\begin{equation}
H^D_{\text{p}}=\hbar\omega_D\sum_l\left(b_l^\dagger b_l+\frac{1}
{2}\right) -\hbar\omega_D\sum_lB_{l+1}^\dagger B_l,
\end{equation}
where,
\begin{equation}\label{Eq:15}
g_D = \alpha\left(\frac{2}{M\omega_D^2J}\right)^{1/2} =
\alpha\left(\frac{1}{KJ}\right)^{1/2}.
\end{equation}
$H^D_{\text{p}}$ may be diagonalized by a Bogoluibov transformation
\cite{kit} to yield,
\begin{equation}
H^D_{\text{p}}=\hbar \sum_q \omega_D(q) \beta_q^{\dagger} \beta_q,
\end{equation}
where $\omega_D(q)$ is the dispersive, gapless phonon spectrum,
\begin{equation}
  \omega_D(q) =  \sqrt{2}\omega_D \sin\left(\frac{q}{2}\right),
\end{equation}
for phonons of wavevector $q$.

\begin{figure}[tb]
\begin{center}
\includegraphics[scale=0.6]{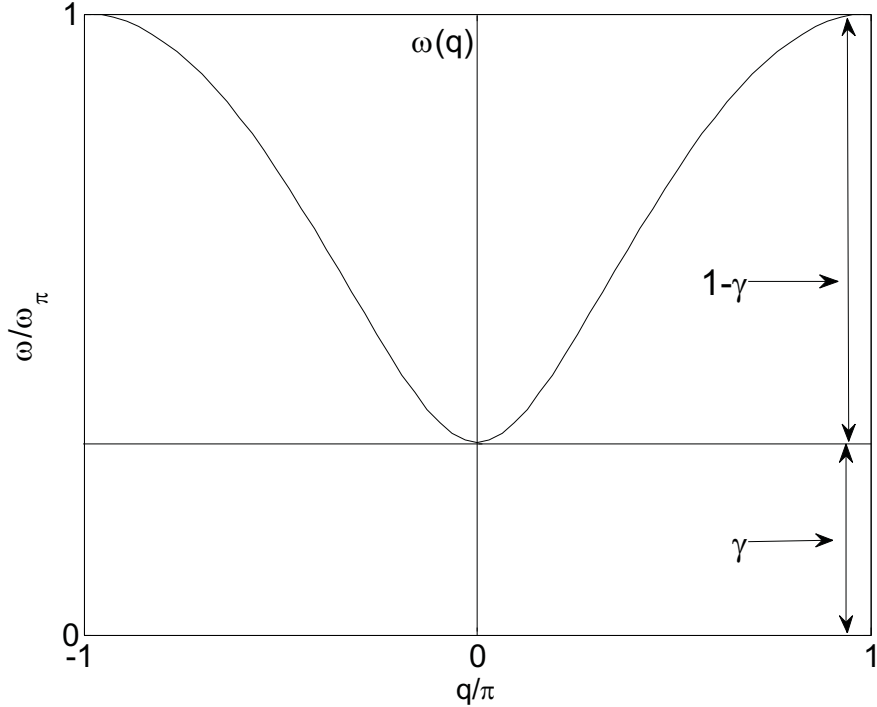}
\end{center}
\caption{Generalized phonon dispersion, $\omega(q)$, defined in Eq.\
(\ref{Eq:20}). $(1-\gamma)\omega_\pi$ is the phonon `band width'
(which vanishes in the Einstein-limit), while $\gamma\omega_\pi$ is
the phonon `mass-gap' (which vanishes in the Debye-limit). The
dispersion parameter, $\gamma$, and the optical phonon frequency,
$\omega_{\pi}$, are model parameters.}\label{disp}
\end{figure}

We now  introduce a generalized  spin-phonon model with a
dispersive, gapped phonon spectrum, via
\begin{equation}\label{Eq:18}
H_{\text{s-p}}=J\sum_l\left[1+g\left(\frac{\hbar\omega_{\pi}}{J}
\right)^{1/2} (B_l-B_{l+1})\right]\textbf{S}_l\cdot \textbf{S}_{l+1}
\end{equation}
and
\begin{equation}\label{Eq:19a}
H_{\text{p}}=\hbar(\omega_E+\omega_D)\sum_l\left(b_l^\dagger
b_l+\frac{1} {2}\right) -\hbar\omega_D\sum_lB_{l+1}^\dagger B_l,
\end{equation}
Again, Eq.\ (\ref{Eq:19a}) may be diagonalized to give,
\begin{equation}\label{Eq:19}
H_{\text{p}}=\hbar \sum_q \omega(q) \beta_q^{\dagger} \beta_q +
\textrm{ constant},
\end{equation}
where,
\begin{equation}\label{Eq:20}
  \omega(q) = (\omega_E+\omega_D)\left( 1- \left(\frac{\omega_D}{\omega_E + \omega_D}\right) \cos q\right)^{1/2},
\end{equation}
is the generalized phonon dispersion, as shown in Fig.\ \ref{disp}.

The $q=0$ phonon gap frequency is,
\begin{equation}\label{Eq:101}
  \omega(q=0) \equiv \omega_0 = \left(\omega_E(\omega_E+\omega_D)\right)^{1/2}
\end{equation}
and the $q=\pi$ optical phonon frequency is,
\begin{equation}\label{Eq:102}
  \omega(q=\pi) \equiv \omega_{\pi} =
  \left((2\omega_E+\omega_D)(\omega_E+\omega_D)\right)^{1/2}.
\end{equation}
We now define the dispersion parameter $\gamma$ as,
\begin{equation}\label{Eq:103}
 \gamma = \omega_0/\omega_{\pi} .
\end{equation}
$\gamma$ is a mathematical device that interpolates the generalized
model between the Einstein ($\gamma = 1$) and Debye ($\gamma = 0$)
limits for a fixed value of the $q=\pi$ phonon frequency,
$\omega_{\pi}$. The dimensionless spin-phonon coupling, $g$,  as
well as $\omega_{\pi}/J$ and $\gamma$ are the independent parameters
in this model. $\omega_E$ and $\omega_D$, on the other hand, are
determined by Eq.\ (\ref{Eq:101}), (\ref{Eq:102}), and
(\ref{Eq:103}).

The generalized model can be mapped onto the  Einstein and Debye
models by the observation that in the Einstein limit,
\begin{eqnarray}\label{Eq:21}
\omega_{\pi} &=& \omega_E \equiv {\omega_b} = \sqrt{K/M};\\
\nonumber g &=& g_E,
\end{eqnarray}
while in the Debye limit,
\begin{eqnarray}\label{Eq:22}
\omega_{\pi} &=& \sqrt{2}\omega_D \equiv 2{\omega_b};\\
\nonumber g &=& g_D/2^{1/4}.
\end{eqnarray}

The introduction of a generalized phonon Hamiltonian avoids the
problems associated with hydrodynamic modes and places a criterion
on the reliability of the gap-crossing characterization of the
critical coupling (as described in Section \ref{gapc}). Starting
from the Heisenberg-SP Hamiltonian in the Einstein limit
($\gamma=1$), the effect of dispersive lattice fluctuations can be
investigated via a variation of $\gamma $. The Debye limit is then
found via an extrapolation of $\gamma \rightarrow 0$.

The model is solved using the density matrix renormalization group
(DMRG) method \cite{white} with periodic boundary conditions
throughout. Our implementation of the DMRG method, including a
description of  the adaptation of the spin-phonon basis and
convergence, is given in the Appendix.

\section{Results and Discussion}\label{Se:3}

\subsection{Gap-crossing}\label{gapc}

\begin{figure}[tb]
\begin{center}
\includegraphics[scale=0.7]{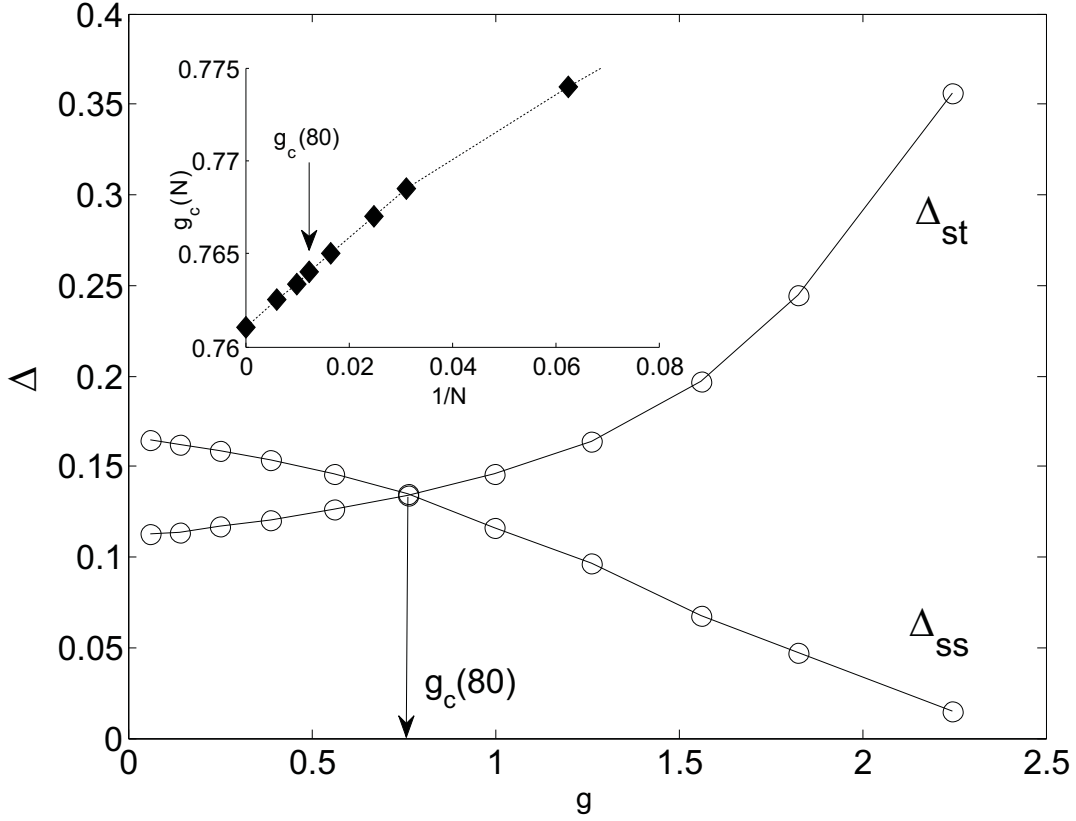}
\caption{Gap-crossing construction for the $\gamma=1$ (Einstein)
Heisenberg-SP model for $N=80$.
 Inset: The infinite-chain critical coupling $g_c^\infty$ is determined by extrapolation.}\label{gapcross}
\end{center}
\end{figure}

For the Einstein model with a non-vanishing value of $\omega_E$ the
critical spin-phonon coupling, $g_{\text c}$, may be determined using
the
gap-crossing method of Okamoto and Nomura \cite{okamoto}, as shown in Fig.\ \ref{gapcross} for an 80-site chain. If the
$N$-site system has quasi-long-range N\'eel order for $0\le g\le
g_{\text c}(N)$, the lowest excitation is a triplet state, i.e.\ $
\Delta_{\text{st}}<\Delta_{\text{ss}}$ and $\lim_{N\to
\infty}\Delta_{\text{st}}=\lim_{N\to \infty}\Delta_{\text{ss}}=0$, where
$\Delta_{\text{st}}$ and $\Delta_{\text{ss}}$ are the triplet and
singlet gaps,
respectively. Conversely, for $g>g_{\text c}(N)$, the system is
dimerized
with a doubly-degenerate singlet GS in the asymptotic limit
(corresponding to the translationally equivalent `A' and `B'
phases), while the lowest energy triplet excitation is gapped.
However, for finite systems the two equivalent dimerization phases
mix via quantum tunneling, and now $\Delta_{\text{ss}}<
\Delta_{\text{st}}$, with
$\lim_{N\to \infty}\Delta_{\text{ss}}=0$ and $\lim_{N\to \infty}
\Delta_{\text{st}}
 >0$.  The gap-crossing condition $\Delta_{\text{st}}=
\Delta_{\text{ss}}$ therefore defines
the finite-lattice crossover coupling $g_{\text c}(N)$.

\begin{figure}[tb]
\begin{center}
\includegraphics[scale=0.6]{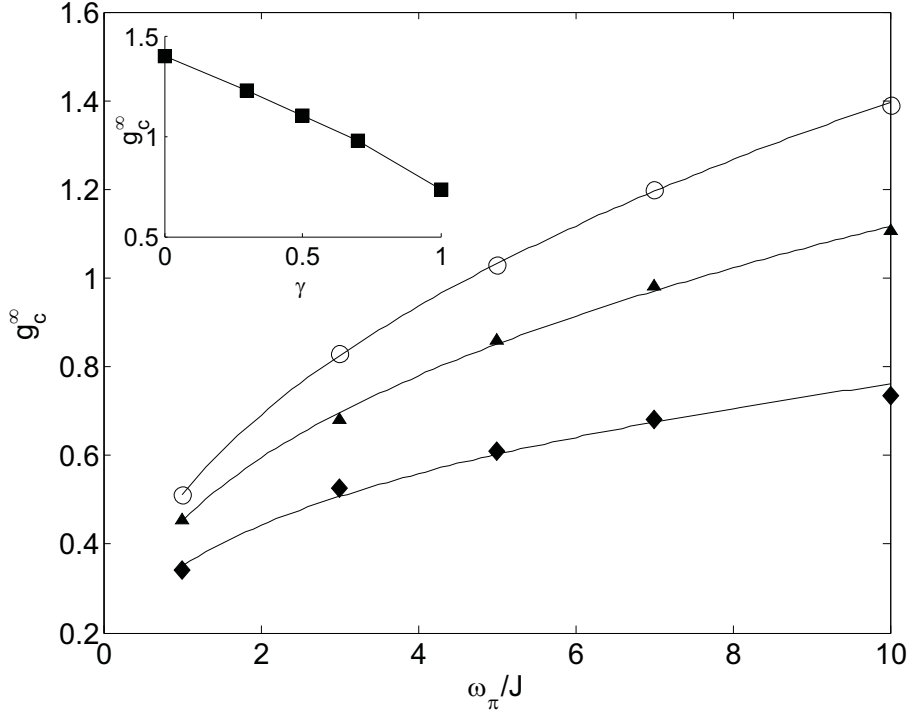}
\end{center}
\caption{Phase diagram in the $g_{\text c}^\infty$-$\omega_\pi$
plane for the infinite  Heisenberg-SP chain for $\gamma=1$
(diamonds) and $\gamma=0.5$ (triangles); extrapolation to $\gamma=0$
generates the Debye-limit (open circles). Inset: variation of
$g_{\text c}^\infty$ with $\gamma$ for the antiadiabatic limit,
$\omega_{\pi}/J=10$.}\label{phaseall}
\end{figure}

For the Debye model, however, the gap-crossing method fails because
of the  $q \to 0$ phonons that form a gapless vibronic progression
with the groundstate. The hybrid spectrum (shown in Fig.\
\ref{disp}) allows us to extrapolate from the pure Einstein limit to
the Debye limit, as the lowest vibronic excitation is necessarily
$\gamma \omega_{\pi}$. Provided that
$\Delta_{\text{ss}}<\omega(q=0)\equiv \gamma\omega_\pi$, the gap
crossover method unambiguously determines the nature of the GS.  We
can confidently investigate Eq.\ (\ref{ham}) for ($0.1\le\gamma\le
1$) with $\omega_\pi/J\in[1,10]$, thereby determining $g_{\text
c}(N,\gamma)$. A polynomial extrapolation of $1/N \to 0$ generates
the bulk-limit critical coupling $g_{\text c}^\infty$ for a given
$\gamma$ (as illustrated in Fig.\ \ref{gapcross}). A subsequent
polynomial extrapolation determines the $\gamma=0$ (Debye) limit.  A
phase diagram for the Heisenberg-SP chain found in this way is shown
in Fig. \ref{phaseall}. Notice that for a fixed $\omega_{\pi}$ the
critical coupling is larger for the Debye model than for the
Einstein model, showing that the quantum fluctuations from the $q <
\pi$ phonons  (as well as the $q = \pi$ phonon) destablize the
Peierls state.

Following Caron and Moukouri \cite{caron} we tentatively propose a
general power-law  for the Heisenberg-SP model, relating the
bulk-limit critical coupling  to $\omega_\pi$ for a given $\gamma$,
\begin{equation}\label{Eq:26a}
g_{\text
c}^\infty(\omega_{\pi},\gamma)=\beta(\gamma)\omega_\pi^{\eta(\gamma)}.
\end{equation}
The infinite-chain values of $\beta$ and $\eta$, and $g_{\text
c}^{\infty} $ for $\omega_{\pi}/J = 10$ are given in Table \ref{nn}.
We find a non-zero critical coupling for all phonon regimes
$\gamma$, with the absolute value of $g_{\text c}^ \infty$
increasing as $\gamma\to 0$, as shown in the inset of Fig.\
\ref{phaseall}.

\begin{table}[h!]
\begin{center}
    \begin{tabular}{| l | c | c || c | c|}
      \hline
      $\gamma$ & $\beta$ & $\eta$ & $g_{\text c}^{\infty}$\\ \hline
      0 (Debye) & 0.511 & 0.437 & 1.397\\
      0.5 & 0.452 & 0.392 & 1.103\\
      1  (Einstein) & 0.350 & 0.337 & 0.761\\
           \hline
    \end{tabular}
\end{center}
\caption{Gap-crossing determined bulk-limit values of
$\beta(\gamma)$ and $\eta(\gamma)$ (defined by Eq.\ (\ref{Eq:26a})),
and $g_{\text c}^{\infty}$ for $\omega_{\pi}/J = 10$. The Debye
limit ($\gamma = 0$) is obtained by extrapolation of $\gamma \to 0$.
}\label{nn}
\end{table}

\subsection{Finite-size scaling}\label{fssg}

\begin{figure}[tb]
\begin{center}
\includegraphics[scale=0.6]{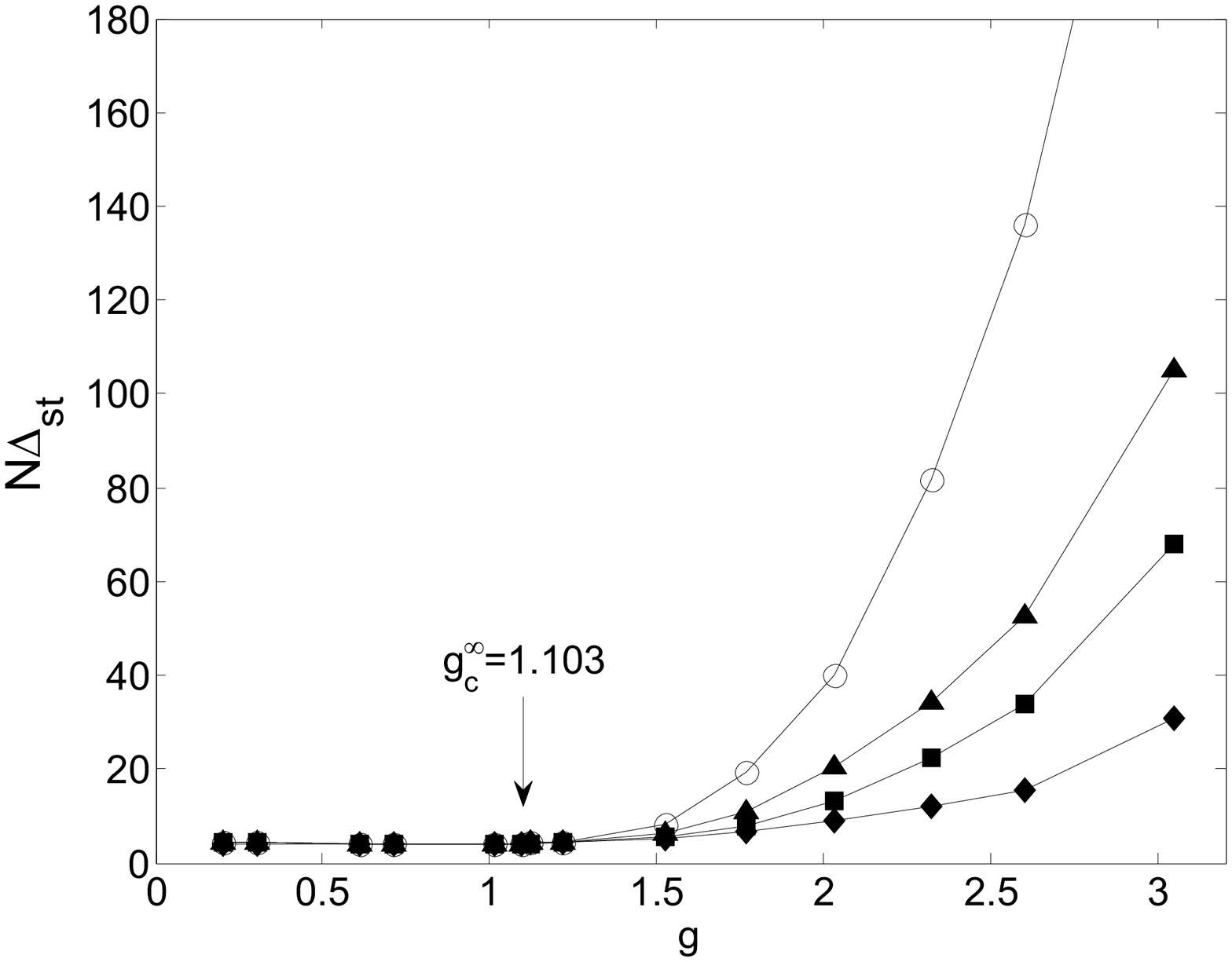}
\end{center}
\caption{$N\Delta_{\text{st}}(N)$ versus the spin-phonon coupling,
$g $, for the $\gamma=0.5$  Heisenberg-SP  model for $N=$ $16$
(diamonds), $40$ (squares), $80$ (triangles), and $160$ (open
circles) for $\omega_\pi/J=10$.  The curves converge at $g_{\text
c}^\infty$ (the value shown is obtained via
gap-crossing).}\label{finiteE5}
\end{figure}

\begin{figure}[tb]
\begin{center}
\includegraphics[scale=0.6]{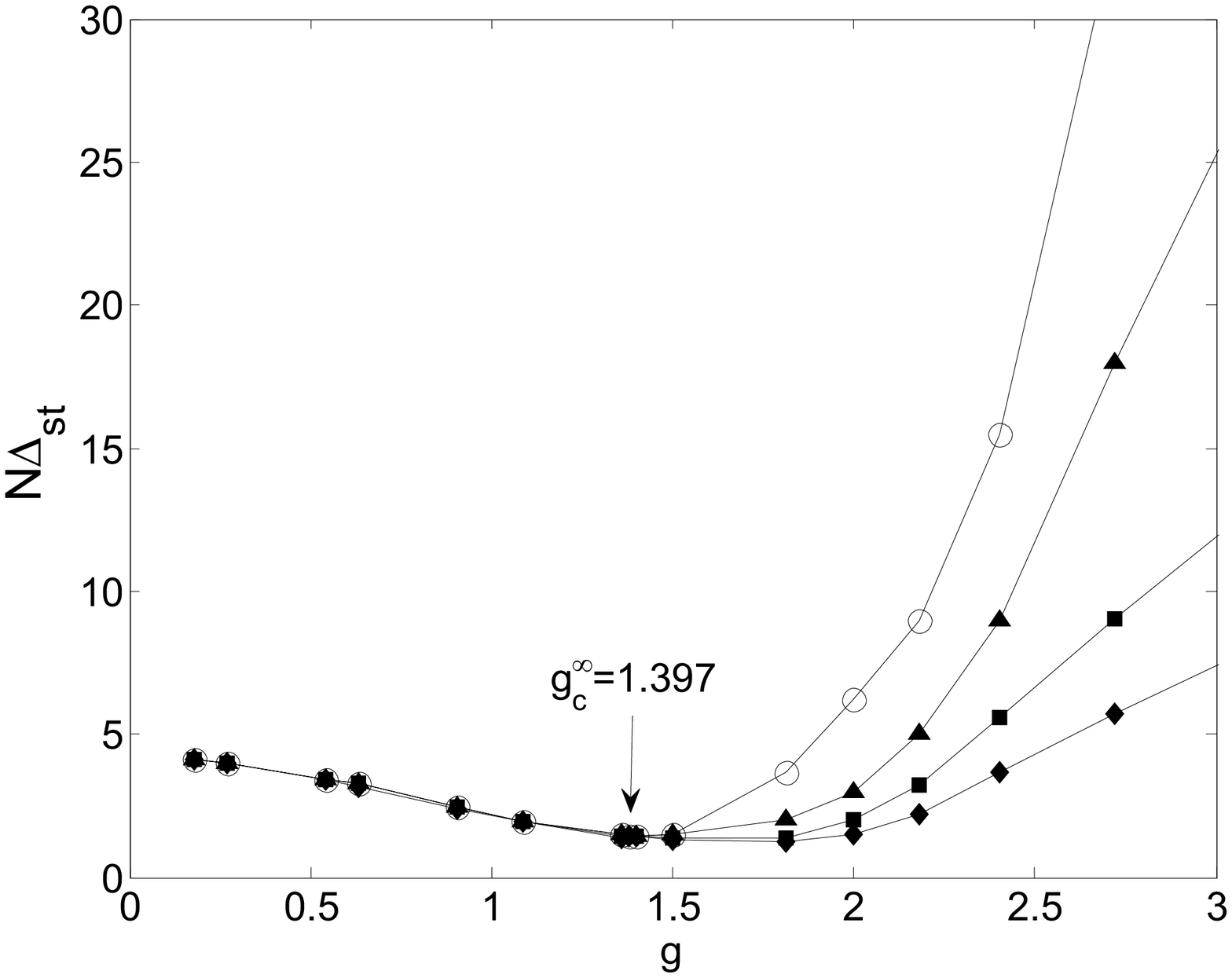}
\end{center}
\caption{$N\Delta_{\text{st}}(N)$ versus the spin-phonon coupling,
$g $, for the $\gamma=0$ (Debye) Heisenberg-SP model  for $N=$ $16$
(diamonds), $40$ (squares), $80$ (triangles), and $160$ (open
circles) for $\omega_\pi/J=10$.  The curves converge at $g_{\text
c}$ (the value shown is obtained via gap-crossing).} \label{FSSE1}
\end{figure}

In order to ascertain the analytic behavior of the spin gap from the
numerical data it is necessary to account for finite-size effects.
We assume that the (singlet-triplet) gap
$\Delta_N\equiv\Delta_{\text{st}}$ for a finite system of $N$ sites
obeys the finite-size scaling hypothesis \cite{fisher,barber}
\begin{equation}\label{fss}
\Delta_N=\frac{1}{N}F(N\Delta_\infty),
\end{equation}
with $\Delta_\infty$ the spin-gap in the bulk limit.  Recalling that
$g_{\text c}^\infty \equiv\lim_{N\to\infty}g_{\text c}(N)$, it follows
that
$\Delta_\infty(g_{\text c}^\infty)=0$ and so curves of $N\Delta_N$
versus
$g$ are expected to coincide at the critical point  where the
bulk-limit spin-gap vanishes, as confirmed in Fig. \ref{finiteE5}
and Fig. \ref{FSSE1}.

The finite-size scaling method is more robust than the gap-crossing
approach, being  applicable to the SP Hamiltonian for all values of
$\gamma$.  On the other hand, its use as a quantitative method is
limited by the accuracy with which plots may be fitted to Eq.
(\ref{fss}).  In practice, plots of $N\Delta_{\text{st}}(N)$ versus
$g$ become progressively more kinked about the critical point as
$\gamma\to 0$.  Nevertheless, we find $F$ to be well approximated by
a rational function and the resulting $g_{\text c}^\infty(\gamma)$
to be in accord with the predictions of the gap-crossover method.

\subsection{Berezinskii-Kosterlitz-Thouless transition}\label{bkt}

\begin{figure}[tb]
\begin{center}
\includegraphics[scale=0.6]{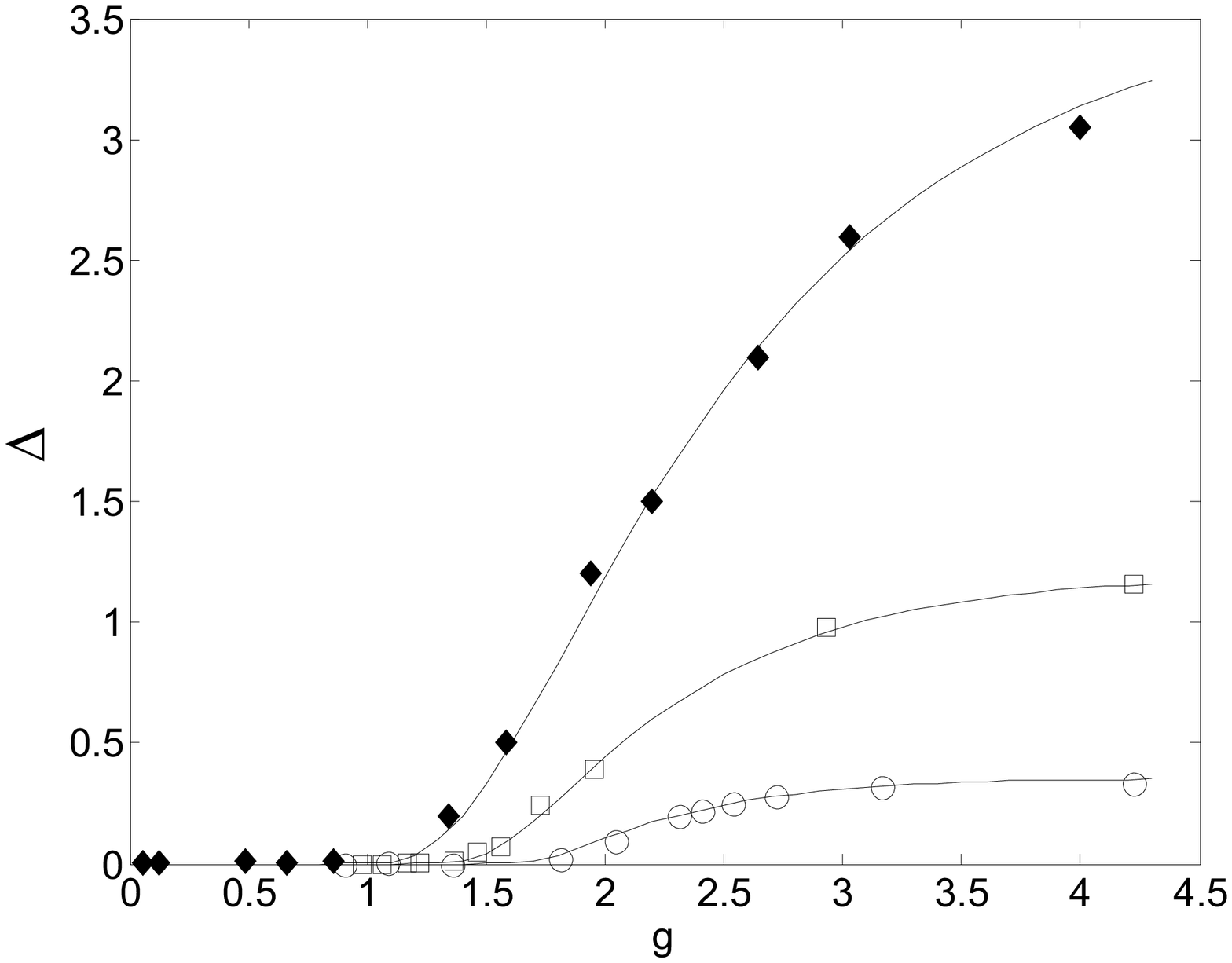}
\end{center}
\caption{Bulk-limit singlet-triplet gap, $\Delta$, as a function of
the spin-phonon coupling, $g$,  with $\gamma=1$ (Einstein)
[diamonds] $\gamma=0.5$ [open squares] and $\gamma=0$ (Debye) [open
circles] for $\omega_\pi/J=10$.  Plots are fitted to the BKT form
(Eq. (\ref{bx})).}\label{KT15}
\end{figure}

For a BKT transition the spin-gap
$\Delta\equiv\lim_{N\to\infty}\Delta_{\text{st}}$ is expected to
exhibit an essential singularity at $g_{\text c}^\infty$ with plots
of $ \Delta_{\text{st}}$ versus $g$ for $N\to\infty$ found to be
well fitted by the Baxter form \cite{baxter}  (as shown in Fig.
\ref{KT15}),
\begin{equation}\label{bx}
\Delta\sim af(g)\exp(-b[f(g)]^2)
\end{equation}
where\cite{bursill},
\begin{equation}\label{bx1}
f(g)\equiv(g-g_{\text c}^\infty)^{-1/2}.
\end{equation}

Extrapolating $\Delta_{\text{st}}(N)$ for $1/N \to 0$ generates
$\Delta$ for a given $\gamma$ and it is  possible, in principle, to
distinguish dimerized from spin-fluid GSs by examining the scaling
behavior of $\Delta_{\text{st}}(N)$, which tends to zero in the
bulk- limit for the spin fluid and to a non-zero $\Delta$ for the
gapped phase. However, not only must three parameters ($a$, $b$, and
$g_{\text c}^ \infty$) be obtained from a non-linear fit (shown in
Table \ref{kttab}), but there is considerable difficulty in
determining $\Delta$ accurately near the critical point: the
spin-gap is extremely small even for values of $g$ substantially
higher than $g_{\text c}^\infty$ due to the essential singularity in
Eq.\ (\ref{bx}).  Determining such small gaps from finite-size
scaling is highly problematic with very large lattices required to
observe the crossover from the initial algebraic scaling (in the
critical regime) to exponential scaling (for gapped systems).  Hence
the gap-crossover method is expected to be substantially more
accurate than a fitting procedure for the determination of the
critical coupling, the latter tending to overestimate $g_{\text
c}^\infty$ (see ref\cite{bursill}), as confirmed by a comparison of
Tables \ref{nn} and \ref{kttab}.

\begin{table}[h!]
\begin{center}
    \begin{tabular}{| l | c | c | c | c |}
      \hline
      $\gamma$ & $a$ & $b$ & $g_{\text c}^\infty$\\ \hline
0 & 1.014 & 1.505 & 1.422\\
       0.5 & 4.110 & 2.101 & 1.120\\
      1 & 14.206 & 3.042 & 0.731\\
      \hline
    \end{tabular}
\end{center}
\caption{Baxter-equation parameters obtained by fits to Eq.\
(\ref{bx}) for $\omega_{\pi}/J = 10$.}\label{kttab}
\end{table}

\subsection{Correlation functions and order parameters}\label{dop}

The $q = \pi$ structure factor, $S(q)$, of the bond-order
auto-correlation function can be used to determine  the phase
transition. $S(q)$ is defined by,
\begin{equation}\label{sf1}
S(q)=\sum_{m}\exp(iqm)\langle C(m)\rangle
\end{equation}
where
\begin{equation}
C(m)=\frac{1}{N}\sum_l(O_l-\langle O_l\rangle)(O_{l+m}-\langle
O_{l+m}\rangle).
\end{equation}
The `bond order' operator, $O_l$, is given by
\[ O_l = \left\{ \begin{array}{ll}
          S_l^z S_{l+1}^z & \mbox{spin-spin};\\
          (q_l-q_{l+1}) & \mbox{phonon displacement}.\end{array}
\right. \]

\begin{figure}[tb]
\begin{center}
\includegraphics[scale=0.6]{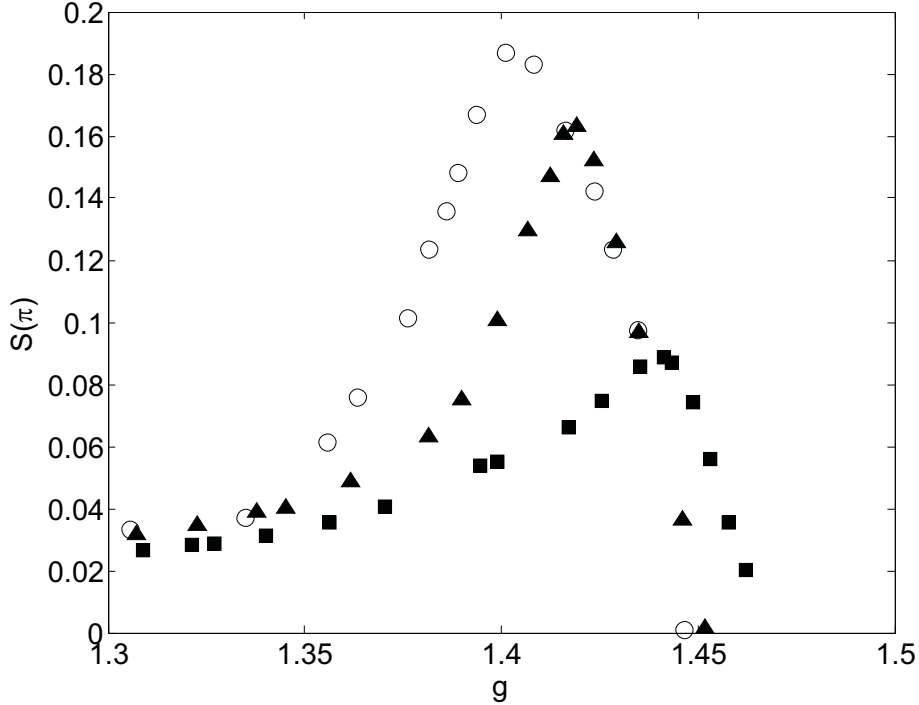}
\end{center}
\caption{Structure factor for the spin-spin bond order
auto-correlation function at $q=\pi$ for the $\gamma=0$ (Debye)
model with $\omega_\pi/J=10$ for $N$=20 (squares), $N$=40
(triangles), and $N$=80 (open circles).}\label{sf}
\end{figure}

The transition to a dimerized state is marked by the development of
a staggered kinetic energy modulation and quasi long-range-order in
the bond order. It is signalled by a divergent peak in $S(q=\pi)$ at
the critical coupling in the asymptotic limit, as shown in Fig.\
\ref{sf}. (We note, however, that the structure factor associated
with the phonon displacement bond order auto-correlator,
$O_l=(q_l-q_{l+1})$, fails to resolve the transition, instead
increasing monotonically with $g$ for all $\gamma$.)

Following \cite{barford1,barford2} we also consider staggered
spin-spin, $m_{\text s}$, and phonon displacement, $m_{\text p}$,
order parameters,
\begin{equation}
m_{\text s}=\frac{1}{N}\sum_l(-1)^l\langle S^z_l S^z_{l +1}\rangle
\end{equation}
and
\begin{equation}
m_{\text p}=\frac{1}{N}\sum_l(-1)^l\langle B_{l+1}-B_l \rangle.
\end{equation}
For linear chains under OBC, the end sites break the energetic
degeneracy between the otherwise  equivalent $\vert A\rangle$ and
$\vert B\rangle$ states, which are related by a translation of one
repeat unit.  Physically, however, PBC are strongly preferable to
OBC as boundary effects are eliminated and finite-size
extrapolations can be performed for much smaller $N$.  In addition,
greater accuracy can be obtained by investigating cyclic chains,
although it is necessary to explicitly break the degeneracy between
the `A' and `B' phases through inclusion of a symmetry-breaking term
$H'$ in Eq.\ (\ref{ham}),
\begin{equation}\label{Eq:30}
H'=\rho\sum_l(-1)^l\langle \textbf{S}_l\cdot\textbf{S}_{l+1}\rangle,
\end{equation}
and extrapolating $\rho \to 0$.

Our results indicate that
 both
$m_{\text s}$ and $m_{\text p}$ scale to zero as $g\to\infty$,
suggesting $g_{\text c}^\infty=0^+$ for all $\gamma$, accounting for
the earlier findings of\cite{barford1}. These predictions are
incomplete deviation to the other positive signatures of a phase
transition for $g>0$. We attribute this discrepancy to the action of
the perturbation $H'$ (Eq.\ (\ref{Eq:30})) on the fixed-point
behavior
 of the Heisenberg-SP Hamiltonian for an insufficiently small
perturbation, and thus conclude that the staggered order parameters
must be treated with caution when determining the phase transition.

\subsection{Quantum bipartite entanglement}\label{bipent}

It has recently been conjectured that quantum entanglement plays an
important role in the quantum phase transitions (QPT) of interacting
quantum lattices.  At the critical point---as in a conventional
thermal phase transition---long-range  correlations pervade the
system.  However, because the system is at $T=0$ (and assuming no
ground-state degeneracy) the GS is necessarily a pure state.  It
follows, then, that the onset of (long-range) correlations -- being
the principal experimental signature of a QPT -- is due to
entanglement in the GS on all length scales.

For an $N$-site lattice, bipartite entanglement is quantified
through the von Neumann entropy \cite{vne},
\begin{equation}\label{vn}
S_{L}=-\mbox{Tr}_{\bar S}\rho_S(L)\log_2\rho_S(L)=-\sum_\alpha\nu_\alpha
\log_2\nu_\alpha,
\end{equation}
where $\rho_S(L)$ is the reduced-density matrix of an $L$-site block
(typically coupled to an $L$-site environment $\bar S$ such that $2L=N
$) and
the $\nu_ \alpha$ are the eigenvalues of $\rho_S(L)$.  It is clear
from Eq. (\ref{vn}) that a slow decay of the reduced density-matrix
eigenvalues corresponds to a large block entropy.  Provided the
entanglement is not too great and the $\nu_\alpha$ decay rapidly, a
matrix-product state is then a good approximation to the GS
\cite{scholl}.  We note here the utility of the DMRG prescription in
determining $S_L\equiv S_{N/2}$ \cite{noack}.

Wu \emph{et al.} \cite{wu} argued, quite generally, that QPTs are
signalled by a discontinuity in some entanglement measure of the
infinite quantum system.  For finite one-dimensional gapped systems
the noncritical entanglement is characterized  by the saturation of
the von Neumann entropy with increasing $L$: the entropy of
entanglement either (i) vanishes for all $L$ or (ii) grows
monotonically with $L$ until it reaches a saturation value for some
block length $L_0$ \cite{vidal}.  \emph{Noncritical} entanglement in
the GS corresponds, thus, to a weak, \emph{semi-local}
\cite{entang1} form of entanglement driven by the appearance of a
length scale $L_0$ due to, e.g, a mass gap in the Hamiltonian.  For
any $L$, the reduced-density matrix $\rho_S(L)$ is effectively
supported on just a small, bounded subspace of the $L$-spin Hilbert
space.  \emph{Critical} models, on the other hand, are expected to
exhibit logarithmic divergence in $S_L$ at large $L$:
$S_L=k\log_2L+\mbox{constant}$.

\begin{figure}[tb]
\begin{center}
\includegraphics[scale=0.6]{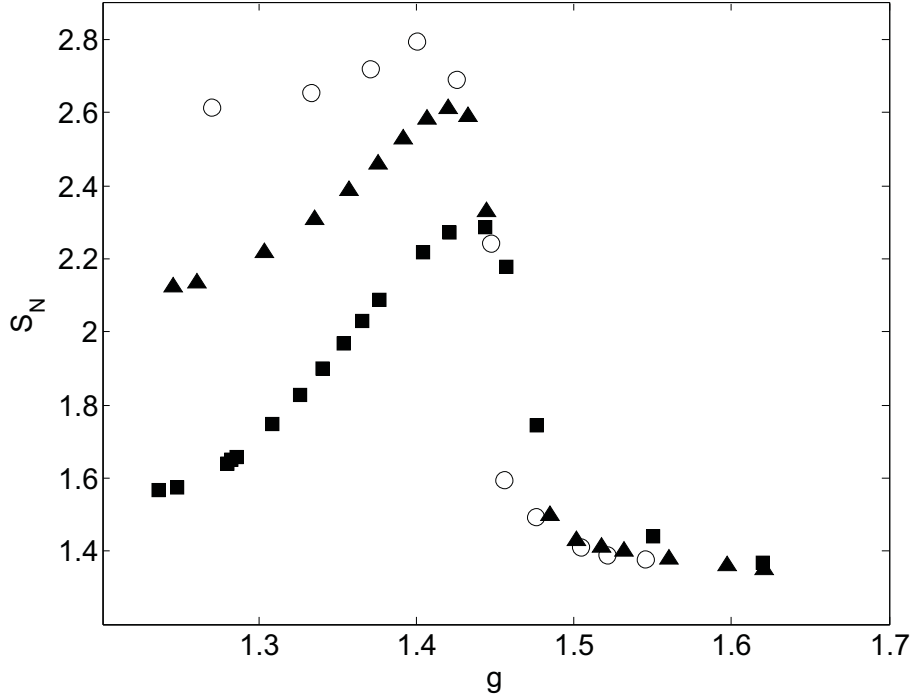}
\end{center}
\caption{Von Neumann entropy, $S_L$, for the $\gamma=0$
  (Debye) model with $\omega_\pi/J=10$ for lattice sizes $N=20$
(squares), $40$ (triangles), and $80$ (open circles);
$L=N/2$.}\label{vonE5}
\end{figure}

For a given total system size $N$ and phonon dispersion $\gamma$,
the block entropy is found to be  maximal for a non-zero spin-phonon
coupling $g_{\text c}(N)$, close to the corresponding gap-crossing and
bond
order structure factor values (as shown in Table \ref{consisttab}).
As shown in Fig.\ \ref{vonE5}, in the critical regime, $g<g_{\text c}
(N)$,
the block entropy is indeed found to scale logarithmically with
system-block length, while in the gapped phase, $g>g_{\text c}(N)$, it
is
characterized by the emergence of a saturation length scale $L_0$
that varies with $\gamma$.  These findings are in agreement with
\cite{vidal} and consistent with the observation that the transition
belongs to the BKT universality class\cite{kt}.

\begin{table}[h!]
\begin{center}
    \begin{tabular}{| l | c | c | c | c |}
      \hline
      $N$ & $g_{\text c}^{\text{gap}}$ & $g_{\text c}^{\text{SF}}$ &
$g_{\text c}^{\text{vN}}$\\ \hline
      20 & 1.441 & 1.441 & 1.443\\
      40 & 1.419 & 1.420 & 1.422\\
      80 & 1.400 & 1.402 & 1.403\\
      \hline
    \end{tabular}
\end{center}
\caption{Consistency of the various probes of the transition:
critical spin-phonon couplings determined by  gap-crossing (gap), $q
= \pi$ structure factor of the bond order auto-correlation function
(SF), and von Neumann entropy (vN) for $N=20$, $40$, and $80$ sites.
$\gamma = 0$ (Debye) and $\omega_\pi/J=10$.}\label{consisttab}
\end{table}

\subsection{Phase diagram}

To conclude this section we discuss the phase diagram of the
Heisenberg spin-Peierls model. Fig.\ \ref{phaseall} shows the phase
diagram as a function of the model parameters $g$ and the $q= \pi$
phonon gap, $\omega_{\pi}$, as defined in Eq.\ (\ref{Eq:18}) and
Eq.\ (\ref{Eq:19}). Evidently, for a fixed value of $\omega_{\pi}$
the spin-Peierls state is less stable to dispersive, gapless quantum
lattice fluctuations than to gapped, non-dispersive fluctuations,
implying that the $q<\pi$ phonons  also destablize the Peierls
state.

\begin{figure}[tb]
\begin{center}
\includegraphics[scale=0.6]{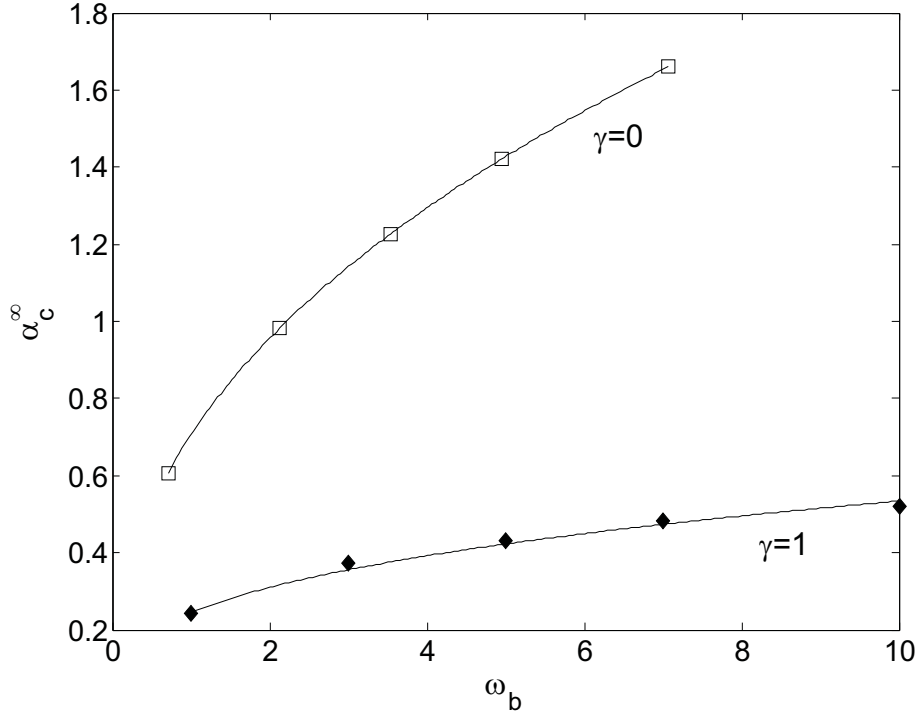}
\end{center}
\caption{Phase diagram in the $\alpha_c^\infty$-${\omega_b}$ plane
for the infinite Heisenberg-SP chain for $\gamma=1$ (Einstein),
diamonds; and $\gamma=0$ (Debye), squares.}\label{phase2}
\end{figure}

It is also instructive, however, to plot the phase diagram as a
function of the \emph{physical} parameters $\alpha$ and ${\omega_b}
= \sqrt{K/M}$, as defined in Eq.\ (\ref{spph}), Eq.\ (\ref{ein}) and
Eq.\ (\ref{deb}). The mapping between model and physical parameters
is achieved via Eq.\ (\ref{Eq:11}), Eq.\ (\ref{Eq:15}),  Eq.\
(\ref{Eq:21}), and  Eq.\ (\ref{Eq:22}) (and setting $K=1$). Since
$\omega_{\pi} = {\omega_b}$ for the Einstein model, whereas
$\omega_{\pi} = 2{\omega_b}$ for the Debye model, the Debye model is
further into the antiadiabatic regime for a fixed value of
${\omega_b}$. We also note that for a given model electron-phonon
coupling parameter, $g$, the physical electron-phonon coupling
parameter, $\alpha$, is larger in the Debye model than the Einstein
model (see Eq.\ (\ref{Eq:11}) and Eq.\ (\ref{Eq:15})). Consequently,
we expect the dimerized phase to be less robust to quantum
fluctuations in the Debye model for fixed values of ${\omega_b}$ and
$\alpha$, as confirmed by Fig.\ \ref{phase2}\cite{footnote1}.

%\begin{figure}[tb]
%\begin{center}
%\includegraphics[scale=0.35]{phase1RSF.eps}
%\end{center}
%\caption{Phase diagram in the $\alpha_c^\infty-\omega_\pi$ plane for
%the infinite Heisenberg-SP chain for  $\gamma=1$ (diamonds) and
%$\gamma=0$ (open squares).}\label{phase1}
%\end{figure}

\section{Conclusions}

The coupling of spin and lattice degrees of freedom with reduced
dimensionality results in the instability of a one-dimensional
Luttinger liquid towards lattice dimerization and the opening of a
gap at the Fermi surface.  Coupling to the lattice gives rise to a
BKT transition from a spin liquid with gapless spinon excitations to
a dimerized phase characterized by an excitation gap.

For the quantum Heisenberg chain in the antiadiabatic limit
($J/\omega_\pi<<1$), the spin-fluid phase becomes  unstable with
respect to lattice dimerization above a non-zero
spin-phonon-coupling threshold for all phonon gaps, $\gamma
\omega_{\pi}$. This observation holds for $\omega_{\pi}/J\sim 1$.
Increasing the contribution of dispersive phonons to $H_{\text{p}}$
gives rise to an increase in the critical coupling, supporting the
intuition that gapless phonons more readily penetrate the GS (with
the $q<\pi$ phonon modes renormalizing the dispersion at the
Peierls-active modes).  This observation has been corroborated by an
array of independent verifications. The behavior of the Debye model
is \emph{qualitatively} different from the Einstein model, with
different exponents and prefactors for the critical coupling versus
phonon frequency, Eq.\ (\ref{Eq:26a}), and the Baxter expression for
the spin-gap, Eq.\ (\ref{bx}).

We note that staggered order parameters are an unreliable means of
determining the phase transition\cite{barford1}, because of the use
of a symmetry breaking perturbation for PBCs that changes the FP of
the Hamiltonian for insufficiently small perturbations.

Placing these findings in the context of experiment, estimates of
the model parameters for a number of spin-Peierls  compounds are
listed in Table \ref{comptab} (reproduced from \cite{bursill}).
Clearly, the static approximation is not applicable to CuGeO$_3$. In
addition, it is also questionable as to whether the listed organic
spin-Peierls compounds are themselves truly \emph{static lattice}
materials, thereby justifying a dynamical phonon treatment.  The
most physically relevant region of the phonon spectrum, however,
appears to be one of intermediate frequency, dividing the adiabatic
and antiadibatic limits.  Nevertheless, even though $\omega_\pi/J=3$
for CuGeO$_3$, referring to the phase diagram of Fig.\
\ref{phaseall} we note the applicability of the generalized
power-law for small $\omega_{\pi}/J$, and hence the occurrence of a
finite critical coupling in the regime applicable to CuGeO$_3$.

\begin{table}[h!]
\begin{center}
    \begin{tabular}{| l | c | c | c | c |}
      \hline
      Material & $J$ & $\omega_\pi$ & $\Delta$\\ \hline
      CuGeO$_3$ & 100 & 300 & 20\\
      TTFCuS$_4$C$_4$(C$_3$F)$_4$ & 70 & 10* & 20\\
      (MEM)(TCNQ)$_2$ & 50 & 100 & 60\\
      \hline
    \end{tabular}
\end{center}
\caption{Estimates of the antiferromagnetic exchange, $J$,
dimerization phonon frequency, $\omega_\pi$, and spin gap, $\Delta$,
for various SP materials.  All units are in  Kelvins.
*Value deduced by comparison of experimental values of $J$, the
transition temperature, and $\Delta$ to mean-field theoretical
expressions in \cite{kasper}.}\label{comptab}
\end{table}

Examination of intermediate phonon frequencies and their extension
to spinful fermion models with Coulomb repulsion\cite{barford2} is
straightforward and currently in progress.

\appendix

\section{DMRG and \emph{in situ} optimization}

We solve Eq.\ (\ref{Eq:18}) and Eq.\ (\ref{Eq:19a}) using the
real-space density matrix renormalization group (DMRG) method
\cite{white}, with ten oscillator levels per site, typically $\sim
200$ block states and ca.\  $10^6$ superblock states.  Finite
lattice sweeps are performed at target chain lengths under PBC.  The
convergence indicators are shown in Tables \ref{e10}-\ref{conv},
with additional convergence tables in ref\cite{barford1} for the
same model.

The DMRG algorithm typically proceeds through the augmentation of an
$ (L-1)$-site \emph{system} block $S$ by a single site $i_a$.  The
augmented system block $S'$ ($L$ sites) is coupled to an augmented
$L$-site \emph{environment} block $E'$, formed analogously to $S'$.
The system and environment comprise the \emph{superblock} ($2L=N$
sites), whose state vector, $\vert\Psi\rangle$,  is readily obtained
by a suitable diagonalization routine.  By tracing over the degrees
of freedom in $E'$, the reduced-density matrix of $S'$
($\rho_{S'}(L)=\mbox{Tr}_{E'}\vert\Psi\rangle\langle\Psi\vert$) is
obtained; a pre-determined proportion of the largest-eigenvalue
eigenstates of $\rho_{S'}(L)$ is retained, forming the system-block
basis for the next iteration.  The DMRG prescription results in
$\mathcal{O}(N)$-growth of the superblock Hilbert space.

In addition (and prior) to the effective truncation and rotation of
the system-block basis, we employ a single-site optimization at each
DMRG step.  All superblock degrees of freedom, save those belonging
to $i_a$, are traced over and the resulting single-site
reduced-density matrix $\rho_a$ is diagonalized, generating an
optimal single-site basis in the correct physical environment  with
which to augment $S$ \cite{lav,barford1,barford2,jeck,zhang}. This
local Hilbert space adaptation generates single-site bases, the
principal utility of which is the solution of many-body problems
with large numbers of degrees of freedom.  A controlled truncation
of a large Hilbert space therefore allows a small (for our purposes
six-dimensional) optimal basis to be used without significant loss of
accuracy, making it ideally suited to spin-phonon problems, where the
number of phonons is not conserved and the phonon Hilbert space is, in
principle, infinite.

\begin{table}[h!]
\begin{center}
    \begin{tabular}{| l | c | c | c | c |}
      \hline
      $m$ & $E_g/J$ & $n_l$ & $\sigma_{n}$\\ \hline
      2 & -18.746372 & 0.00291 & 0.0468\\
      5 & -18.747323 & 0.00221 & 0.0470\\
      8 & -18.747323 & 0.00221 & 0.0470\\
      10 & -18.747367 & 0.00221 & 0.0470\\
      \hline
    \end{tabular}
\end{center}
\caption{GS energy, $E_g/J$, average phonon occupation number,
$n_l=\langle b_l^\dagger b_l\rangle$, and standard  deviation,
$\sigma_n$, for a 40-site chain with $\gamma=1$ (Einstein),
$\omega_{\pi}/J=10$, and $m$ oscillator levels per site.}\label{e10}
\end{table}

\begin{table}[h!]
\begin{center}
    \begin{tabular}{| l | c | c | c | c |}
      \hline
      $m$ & $E_g/J$ & $n_l$ & $\sigma_{n}$\\ \hline
      2 & -27.198068 & 0.0331 & 0.179\\
      5 & -31.795996 & 0.1477 & 0.437\\
      8 & -31.991408 & 0.1717 & 0.481\\
      10 & -31.991541 & 0.1719 & 0.484\\
      \hline
    \end{tabular}
\end{center}
\caption{GS energy, $E_g/J$, average phonon occupation number,
$n_l$, and standard deviation, $\sigma_n$, for a  40-site chain with
$\gamma=0$ (Debye), $\omega_{\pi}/J=10$, and $m$ oscillator levels
per site.}\label{20sites}
\end{table}

\begin{table}[h!]
\begin{center}
    \begin{tabular}{| l | c | c | c | c |}
      \hline
      $m$ & $E_g/J$ & $n_l$ & $\sigma_{n}$\\ \hline
      2 & -18.882263 & 0.0318 & 0.175\\
      5 & -19.358486 & 0.1368 & 0.420\\
      8 & -19.371279 & 0.1520 & 0.451\\
      10 & -19.371877 & 0.1522 & 0.452\\
      \hline
    \end{tabular}
\end{center}
\caption{GS energy, $E_g/J$, average phonon occupation number,
$n_l$, and standard deviation, $\sigma_n$, for a  40-site chain with
$\gamma=0$ (Debye), $\omega_{\pi}/J=1$, and $m$ oscillator levels
per site.}\label{d1}
\end{table}

\begin{table}[h!]
\begin{center}
    \begin{tabular}{| l | c | c | c | c |}
      \hline
      $\epsilon$ & $E_g/J$ & $M$ & SBHSS\\ \hline
      $10^{-10}$ & -24.266060 & 176 & 14572 \\
      $10^{-11}$ & -24.350806 & 278 & 32328\\
      $10^{-12}$ & -24.377641 & 300 & 60054\\
      $10^{-15}$ & -24.377646 & 350 & 172654\\
      \hline
    \end{tabular}
\end{center}
\caption{GS energy, $E_g/J$, of Heisenberg-SP model as a function of
the density-matrix eigenvalue product  cutoff, $\epsilon$, number of
system block states, $M$, and the superblock Hilbert space size,
(SBHSS) for a 40-site chain with 10 oscillator levels per site and
$\gamma=0.5$.}\label{40sites}
\end{table}

\begin{table}[h!]
\begin{center}
    \begin{tabular}{| l | c | c | c | c |}
      \hline
      $N$ & $\epsilon$ & $\Delta_{\text{ss}}$ & $\Delta_{\text{st}}$\\
\hline
8 & $10^{-12}$ & 0.924728 & 0.513871\\
8 & $10^{-15}$ & 0.924726 & 0.513869\\
8 & $10^{-20}$ & 0.924726 & 0.513869\\\hline

20 & $10^{-12}$ & 0.343794 & 0.215902\\
20 & $10^{-14}$ & 0.343792 & 0.215900\\
20 & $10^{-16}$ & 0.343792 & 0.215899\\\hline

40 & $10^{-12}$ & 0.183655 & 0.113428\\
40 & $10^{-13}$ & 0.183641 & 0.113405\\
40 & $10^{-14}$ & 0.183640 & 0.113403\\
40 & $10^{-15}$ & 0.183640 & 0.113403\\ \hline

160 & $10^{-12}$ & 0.046367 & 0.030776\\
160 & $10^{-13}$ & 0.046163 & 0.030550\\
160 & $10^{-14}$ & 0.046161 & 0.030528\\
160 & $10^{-15}$ & 0.046163 & 0.030526\\
      \hline
    \end{tabular}
\end{center}
\caption{DMRG convergence of the singlet, $\Delta_{\text{ss}}$, and
triplet,
  $\Delta_{\text{st}}$, gaps of the  Heisenberg-SP model for $
\gamma=0$ with
density-matrix eigenvalue product cutoff, $\epsilon$, for various
$N$-site periodic lattices, where $\omega_\pi/J=10$ and
$g=0.4$}\label{conv}
\end{table}

\begin{acknowledgements}
We  thank Professor Fabian Essler  for invaluable discussions.
\end{acknowledgements}


\begin{references}
\bibitem{entang1} A good approximation to $\rho_S(L)$ can be obtained
by diagonalizing the Hamiltonian corresponding to the system block
and only a few extra, neighboring/environment-block spins, e.g.
DMRG.
\end{references}
\end{document}